\documentclass[twocolumn,
amsmath,amssymb,floatfix,10pt,prd,
superscriptaddress,nofootinbib]{revtex4}

\usepackage[parfill]{parskip}    
\usepackage{graphicx}
\usepackage{amssymb}
\usepackage{epstopdf}
\usepackage{color}
\usepackage{float}
\usepackage{amsmath}
\usepackage{orcidlink}
\usepackage{latexsym}
\usepackage{amsmath}
\usepackage{amssymb}
\usepackage{amsfonts}
\usepackage{subcaption}
\usepackage[font=small,labelfont=bf,
   justification=justified,
   format=plain]{caption}

\begin{document}

\title{Unified first law of thermodynamics in Gauss-Bonnet gravity on an FLRW background }

\author{Joel  Saavedra
 \orcidlink{0000-0002-1430-3008}}
\email{joel.saavedra@pucv.cl}
\affiliation{
Instituto de Física, Pontificia Universidad Católica de Valparaíso, Casilla 4950, Valparaíso, Chile.}

\author{Francisco Tello-Ortiz
 \orcidlink{0000-0002-7104-5746}}
\email{francisco.tello@pucv.cl}
\affiliation{
Instituto de Física, Pontificia Universidad Católica de Valparaíso, Casilla 4950, Valparaíso, Chile.}

\begin{abstract}

Employing the thermodynamics unified first law through the thermodynamics-gravity conjecture, in this article, we derive for an FLRW universe the Friedmann equations in the framework of Gauss-Bonnet gravity theory. To do this, we project this generalized first law along the Kodama vector field and along the direction of an orthogonal vector to the Kodama vector. The second Friedmann equation is obtained by projecting on the Kodama vector, while the first is obtained by projecting along the flux on the Cauchy hypersurfaces. This result does not assume a priory temperature and an entropy, so the Clausius relation is not used here. Nevertheless, it is used to obtain the corresponding Gauss-Bonnet entropy. In this way, the validity of the generalized second law of thermodynamics is proved for the Gauss-Bonnet gravity theory.

\end{abstract}

\maketitle

\section{Introduction}

The connection between gravity and thermodynamics, or in other words, the thermodynamics description of gravity and vice versa, was put forward by seminal articles by Bardeen, Carter and Hawking \cite{Bardeen:1973gs}, Bekenstein \cite{Bekenstein:1973ur}, Hawking \cite{Hawking:1975vcx}, Gibbons and Hawking \cite{Gibbons:1976ue} and York \cite{York:1986it}. In this series of works, the black hole (BH) thermodynamics laws were developed, leading to an important interpretation such as the BH temperature as the surface gravity \cite{Hawking:1975vcx, Gibbons:1976ue} and the BH entropy as the surface area \cite{Bekenstein:1973ur}. All the mentioned results were obtained within the framework of General Relativity (GR); however, later on, this analysis was extended to higher dimensional gravitational theories, specifically Lovelock's gravity \cite{Lovelock:1971yv}, by Myers and Simon in \cite{Myers:1988ze}. After all these advances, Jacobson obtained Einstein's field equations from a pure thermodynamics description \cite{Jacobson:1995ab}.

The aforementioned unexpected link gave rise to possible quantum gravitational effects, taking a first step in unifying these two branches: gravity and quantum mechanics. However, an additional crucial point was necessary to proceed further in this direction. In doing that, Hayward, in a series of articles \cite{Hayward:1993wb, Hayward:1994bu, Hayward:1997jp}, derived the thermodynamics unified first law (UFL) for dynamical BHs. In this context, the Killing vector field is replaced by the Kodama vector field \cite{Kodama:1979vn} and event horizons by apparent horizon \cite{Hayward:1997jp, Kodama:1979vn, Faraoni:2011hf}. Interestingly, the Kodama vector field establishes a privileged time direction, and its contraction with Einstein tensor defines a conserved charge \cite{Kodama:1979vn, Hayward:1997jp}.

This dynamical framework allowed us to spread the thermodynamics description to scenarios beyond the BHs arena. Concretely, this methodology has been applied to the cosmological scenario in several gravity theories \cite{Cai:2005ra, Akbar:2006kj, Cai:2006rs, Sheykhi:2010zz, Sebastiani:2023brr}, where one does not have an event horizon, instead one an inner past apparent horizon (for expanding cosmology) or future inner apparent horizon (for contracting cosmology)\cite{Faraoni:2011hf, Faraoni:2015ula, Helou:2015zma}.

In this concern, it has been argued that to obtain the dynamic of the FLRW universe, i.e., Friedmann field equations, one needs to assume from the very beginning  the notion of temperature and entropy \cite{Cai:2005ra, Cai:2006rs}. This is  because the term describing the energy variation in the UFL becomes zero when projecting it along the apparent horizon. For the GR case, in \cite{Helou:2015yqa, Helou:2015zma}, it  was proven that this happens for every constant radius and not only for the apparent horizon. However, this conclusion drastically changes when different vector fields are used to project the UFL along the apparent horizon. For example, when contracting the UFL with the Kodama vector, the Friedmann equations naturally arise without further consideration.

On the other hand, if one follows the approach given in \cite{Cai:2005ra, Cai:2006rs} in the Gauss-Bonnet gravity theory (as was done in \cite{Cai:2005ra}), it is necessary to assume that the apparent horizon is fixed when projecting the UFL along it. This is because the Misner--sharp energy \cite{Misner:1964je} contains non-trivial contribution coming from the gravitational sector \cite{Maeda:2006pm, Maeda:2007uu, Maeda:2008nz}, and these do not vanish unless the apparent horizon is taken to be constant, as previously mentioned. Nevertheless, as the apparent horizon is a dynamic object \cite{Faraoni:2015ula}, considering it a fixed entity is a very strong restriction. Then, in this article, we do not follow the approach used in \cite{Cai:2005ra} to regain Friedmann equations from the UFL. Instead of it, we project the UFL along the apparent horizon using the Kodama vector field. This is possible because the Kodama vector becomes a null vector on the apparent horizon \cite{Kodama:1979vn, Hawking:1975vcx, Faraoni:2015ula}; in this case, the apparent horizon behaves as a null surface. This means that the Kodama vector is orthogonal and tangent to the apparent horizon. In this way, the second Friedmann equation naturally arises without the assumption of Clausius thermodynamics relation. Then, the first Friedmann equation is obtained by projecting the UFL along the direction of a vector field orthogonal to the Kodama vector, consequently hypersurface orthogonal to the apparent horizon surface. The Clausius relation becomes relevant in testing the so-called generalized second law. The Gauss-Bonnet entropy is also obtained by starting from the UFL and identifying the energy-supply contribution with the Clausius relation.

The article is organized as follows: Sect. \ref{sec2} provides a short revision about the apparent horizon description for an FLRW universe and a short review of the Kodama vector field in this background. Sect. \ref{sec3} is devoted to the presentation of the Gauss-Bonnet gravity theory and also presents the main results of this research, that is, the Friedmann equations obtained from the UFL (\ref{3SB}) and the Gauss-Bonnet entropy expression (\ref{3SC}). Finally, Sect. \ref{sec4} provides some remarks about the article.

Throughout the article, the mostly positive signature $\{-,+,+,+\}$ and units where the speed of light $c=1$ are used.

\section{Apparent horizon and Kodama vector field}\label{sec2}

In this section, we briefly review the main and elemental ingredients to study the thermodynamics of the FLRW space-time. To do this, we must define the apparent horizon in describing dynamical solutions \cite{Faraoni:2011hf, Faraoni:2015ula}. Along with this information, determining the proper dynamical vector field is necessary, providing the correct information on the apparent horizon. This vector field is the well-known Kodama vector field \cite{Kodama:1979vn, Hayward:1993wb, Hayward:1994bu, Hayward:1997jp} which is replacing the usual time--like Killing vector field employed in the static situation.  

\subsection{The apparent horizon of the FLRW space--time}

{As stated before, the pertinent entity to describe the thermodynamics of the Universe is the so-called apparent horizon ($AH$) \cite{Faraoni:2011hf, Faraoni:2015ula}. The $AH$ is the surface satisfying 
\begin{equation}\label{EqAH}
    h^{ij}\nabla_{i}R\nabla_{j}R=0, \quad i,j=t,r,
\end{equation}
where $R$ is the areal radius defined as $R(t,r)=a(t)r$, being $a(t)$ the scale factor,  $r$ the co--moving radial distance and $h^{ij}$ the inverse metric of the two--dimensional metric $h_{ij}$ on the $t-r$ plane. These definitions come from the FLRW line element 
\begin{equation}\label{OFLRW}
d s^2=-d t^2+a^2(t)\left(\frac{d r^2}{1-k r^2}+r^2 d \Omega_{(n-2)}^2\right),
\end{equation}
re-expressed by means of the warped product as
\begin{equation}\label{warped}
    ds^{2}=h_{ij}dx^{i} dx^{j}+R^{2}d \Omega_{(n-2)}^2,
\end{equation}
where in the above expressions $k=0,\pm 1$ is the spatial curvature and $d \Omega_{(n-2)}^2$ is the line element of a $(n-2)$--dimensional unit sphere. Therefore, in this way the metric on $t-r$ plane is given by $h_{ij}=\text{diag}\{-1,a^{2}/\left(1-kr^{2}\right)\}$.

Now, the solution $R_{AH}$ of Eq. (\ref{EqAH}) is given by\footnote{The apparent horizon in this context, is a dynamical marginally outer trapped surface (MOTS) with vanishing
expansion \cite{Faraoni:2015ula}.}\cite{Faraoni:2011hf,Faraoni:2015ula}
\begin{equation}\label{AHvalue}
    R_{AH}=\frac{1}{\sqrt{H^{2}+\frac{k}{a^{2}}}},
\end{equation}
with $H=\dot{a}/a$ being the Hubble's parameter. For a flat FLRW, when $k=0$, the expression (\ref{AHvalue}) coincides with the Hubble's radius. From now on, we will keep the spatial curvature $k$ arbitrary.

Although the co-moving patch\footnote{Greeks indexes run over the full space-time components, Latin indexes only run over time and radial components.} $x^{\mu}=\{t,r,\theta_{1}\ldots \theta_{n-2}\}$, with $n-3$ angles ranging over $[0,\pi]$ and $n-2$ ranging over $[0,2\pi)$, is the natural chart for the FLRW metric, it is more instructive to express this model using the Schwarzschild--like coordinate chart $y^{\mu}=\{T,R,\theta_{1}\ldots \theta_{n-2}\}$. One must establish the relation between the co-moving time $t$ and radius $r$ with the new time $T$ and radius $R$ coordinates to do this. These relations are given by \cite{Faraoni:2011hf,Faraoni:2015ula,Nielsen:2005af}  
\begin{equation}\label{time}
    dt=FdT-\beta dR,
\end{equation}
\begin{equation}
    dr=-\frac{HRF}{a}dT+\frac{1}{a}dR,
\end{equation}
where $F=F(T,R)$ is an integration factor in order to assure that $dt$ being an exact differential, $H$ and $a$ implicit functions of the time coordinate $T$ and $\beta$ a function given by
\begin{equation}
    \beta(t,R)=\frac{HR}{1-H^{2}R^{2}-kr^{2}}.
\end{equation}
In this way, the line element (\ref{OFLRW}) acquires the following form
\begin{equation}\label{schmetric}
    \begin{split}
        ds^{2}=-\left(1-\frac{H^{2}R^{2}}{1-kR^{2}/a^{2}}\right)F^{2}dT^{2} &\\  +\frac{dR^{2}}{1-kR^{2}/a^{2}-H^{2}R^{2}}+R^{2}d \Omega_{(n-2)}^2.
    \end{split}
\end{equation}

From (\ref{schmetric}) it is easy to get the $AH$ by solving 
\begin{equation}
    g^{RR}=0,
\end{equation}
for $R$ leading as before to (\ref{AHvalue}).

\subsection{Kodama vector field for FLRW space--time}

As it is well-known, there is no preferred time coordinate in an evolving time-dependent space-time. This is because there is no asymptotically time--like a Killing vector field. Then, to bypass this issue, a divergence-free vector field which is present in any time-dependent spherically symmetric\footnote{For generalizations of the Kodama vector to symmetries other than the spherical one see for example \cite{Tung:2007vq, Kinoshita:2021qsv}.} space-time was introduced in \cite{Kodama:1979vn} and is referred as the Kodama vector field. The Kodama vector field identifies a natural time-like direction and is used to define locally conserved currents \cite{Kodama:1979vn, Hayward:1993wb}. So, for dynamical situations, it is necessary to introduce or replace the objects used in a static/stationary context for those giving the proper description \cite{Hayward:1997jp}, that is
\begin{enumerate}
    \item Killing vector field $\mapsto$ Kodama vector field.
    \item Killing horizon $\mapsto$ apparent horizon.
\end{enumerate}

In general, the Kodama vector field is defined as
\begin{equation}\label{Kodama}
K^{i} \equiv E^{i j} \nabla_{j} R,
\end{equation}
where $E^{ij}$ is the skew-symmetric Levi--Civita tensor defines as
\begin{equation}
    E^{i_{1}\ldots i_{n}}=\frac{\text{sgn}\left(|h|\right)}{\sqrt{|h|}}\varepsilon^{i_{1}\ldots i_{n}},
\end{equation}
where $h$ is the determinant of the metric $h_{ij}$, the sign function $\text{sgn}(h)=(-1)^{q}$, with $q$ the number of negatives in the metric signature\footnote{This definition occurs when the metric has an odd number of negative signs in the signature as in our case.} and $\varepsilon^{i_{1}\ldots i_{n}}$ the standard Levi--Civita symbol defined according our basis coordinate orientation as
\begin{equation}
\varepsilon_{i_1 \ldots i_n}=-\varepsilon^{i_1 \ldots i_n}\left\{\begin{aligned}
+1 &   \text { even permutation, } \\
-1 &  \text { odd permutation, } \\
0 & \text { otherwise. }
\end{aligned}\right.
\end{equation}

On the other hand, if the metric is expressed in Schwarzschild--like coordinate gauge, the components of the Kodama vector (\ref{Kodama}) assume the simple form 
\cite{Kodama:1979vn,Racz:2005pm}
\begin{equation}
    K^{i}=\frac{1}{\sqrt{g_{TT}g_{RR}}}\left(\frac{\partial}{\partial t}\right)^{i}.
\end{equation}
Then, for the line element (\ref{schmetric}) one gets
\begin{equation}\label{kodamasch}
    K=K^{T}\partial_{T}=F\sqrt{1-\frac{k}{a^{2}}R^{2}}\,\partial_{T}.
\end{equation}

Now, from the norm of the Kodama vector $K^{2}=K_{i}K^{i}$ 
\begin{equation}\label{kodamavector}
K^{2}=K_i K^i=-\left(1-H^2 R^2-\frac{k R^2}{a^2}\right)=-\left(1-\frac{R^2}{R_{A H}^2}\right),
\end{equation}
we can infer the following valuable information: the Kodama vector is time--like if $K^{2}<0$, thus $R<R_{{AH}}$, null $K^{2}=0$ if $R=R_{{AH}}$ and space--like if $K^{2}>0$, that is, outside the $AH$ $R>R_{{AH}}$. Thus, the $AH$ behaves as a null surface in this case. Fig. \ref{fig1} shows a schematic representation of the Kodama vector (\ref{kodamasch}). As can be appreciated, in the chart $y^{\mu}=\{T,R,\theta_{1}\ldots \theta_{n-2}\}$ it is hypersurface orthogonal to the 
Cauchy hyper--surface $\Sigma(R,\theta_{1}\ldots \theta_{n-2}): T-T_{0}=0$. This fact is clear from the general hyper--orthogonal condition, which reads: $X_{i}=\Lambda(x^{j})\nabla_{i}\Sigma$, being $\Lambda$ some proportionality factor, which in general will vary from point to point and $\nabla_{i}\Sigma=\partial_{T}$. Besides, the Kodama vector is orthogonal to the two--spheres of symmetry. Interestingly, the modulus of the norm (\ref{kodamavector}) coincides with the space--like slice of marginally trapped tube (MTT) given by\footnote{As we are performing a foliation of the space-time, as shown in Fig. \ref{fig1}, the surface $\chi$ is just a two-sphere of radius $R_{AH}$ embedded in a specific Cauchy hypersurface $\Sigma$. However, if we collect all the constant time slices $\Sigma$ one gets the MTT (purple surface in Fig. \ref{fig1}), equivalently if one considers $\chi(T,\theta_{1}\ldots \theta_{n-2})$ instead of only $\chi(\theta_{1}\ldots \theta_{n-2})$ one obtains the full MTT.}
\begin{equation}\label{sphere}
    \chi(\theta_{1}\ldots \theta_{n-2}): 1-\frac{R^{2}}{R^{2}_{AH}}=0.
\end{equation}
It is worth mentioning that each space--like slice (red circles) defines the MOTS given rise to the $AH$ \cite{Faraoni:2011hf, Faraoni:2015ula}.

In analogy with Killing vectors, the Kodama vector satisfies
\begin{equation}\label{kodamaEQ}
\frac{1}{2} h^{ij} K^k\left(\nabla_k K_i-\nabla_i K_k\right)=\kappa_{\text {Kodama }} K^j,
\end{equation}
where $\nabla_{i}$ is covariant the derivative compatible with the metric $h_{ij}$ ($\nabla_{k}h_{ij}=0$).
The Eq. (\ref{kodamaEQ}) defines the notion of surface gravity for dynamical scenarios. Usually, this is referred to as the Hayward--Kodama (HK) surface gravity \cite{Hayward:1997jp}, which in general has the following expression independent of the gravitational theory at hand
\begin{equation}\label{SG}
\kappa_{\text{HK }}=\frac{1}{2 \sqrt{|h|}} \partial_i\left(\sqrt{|h|} h^{ij} \partial_j R\right).
\end{equation}

Now, we can discuss the thermodynamics of the FLRW space-time (\ref{OFLRW}) through the associated thermodynamics UFL in the context of the higher dimensional GB theory.

\begin{figure}[H]
    \centering
\includegraphics[width=0.5\textwidth]{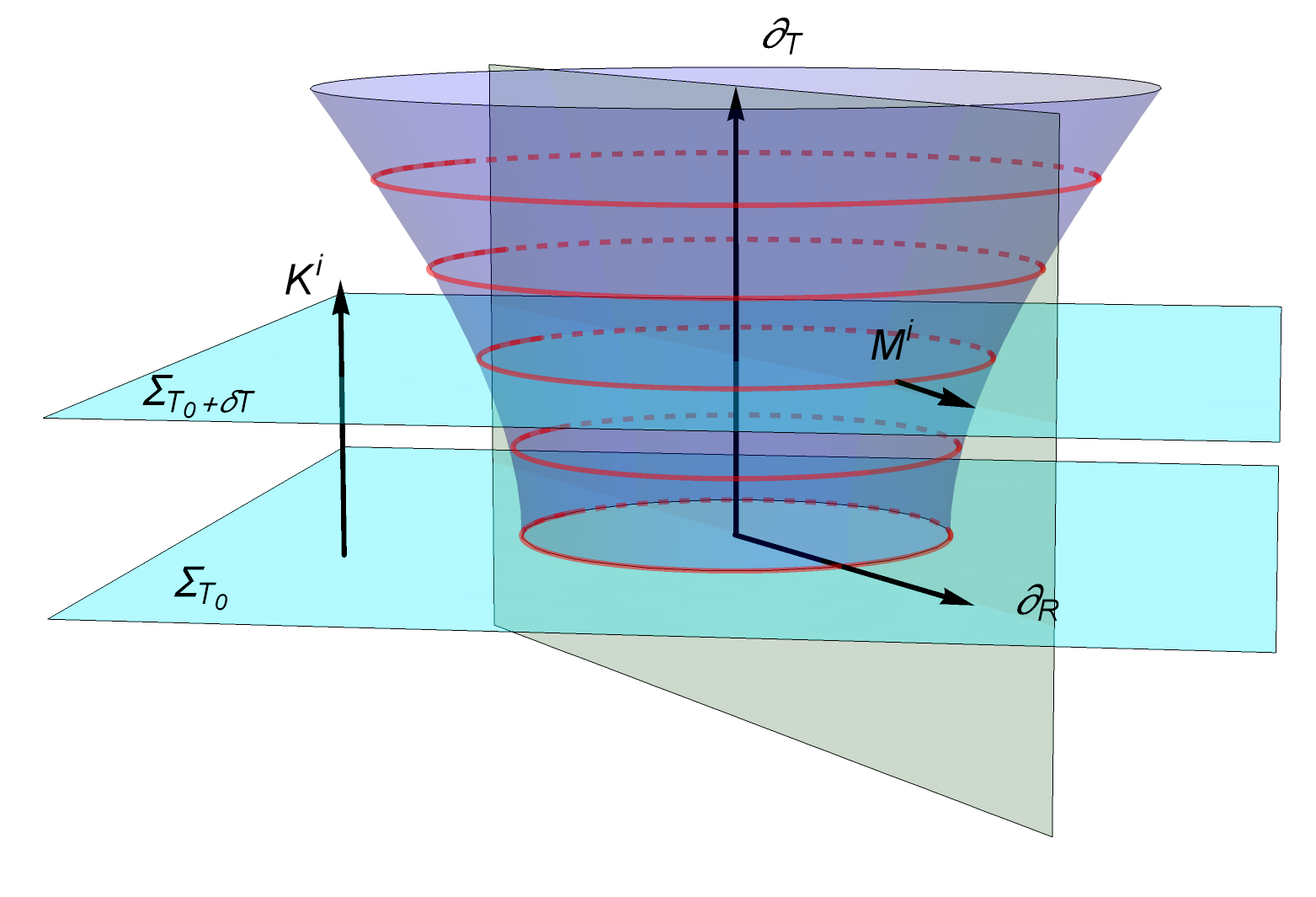}
    \caption{The Kodama vector $K^{i}$ parallel to the time--like vector field $\partial_{T}$ orthogonal to the space--like Cauchy hyper--surfaces $\Sigma=\Sigma_{T_{0}}(R,\theta,\phi):T-T_{0}=0$. 
    The marginally trapped tube (MTT) corresponds to the purple surface. Each space-like surface (red circles) corresponds to marginally outer trapped surfaces (MOTS), giving rise to the apparent horizon ($AH$) \cite{Faraoni:2015ula}, on which the norm of the Kodama vector nullifies. This schematic representation is constructed using the Schwarzschild--like coordinate patch $y^{\mu}=\{T,R,\theta_{1}\ldots \theta_{n-2}\}$. }
    \label{fig1}
\end{figure}

\section{thermodynamics UFL in GB gravity }\label{sec3}

\subsection{Preliminaries of GB gravity}\label{3SA}

The whole action of the $n$--dimensional GB\footnote{This theory is also known as Einstein-Gauss-Bonnet (EGB) gravity; however, we will refer to this theory as GB only.} couple to matter fields, is given by

\begin{equation}\label{fullaction}
S_{GB}=\frac{1}{16\pi G_{n}}\int_{\mathcal{M}} d^{n} x \sqrt{-g}\left[\mathcal{R}+\alpha\mathcal{G}_{\text{GB}}\right] + \text{B.T.} + S_{m},
\end{equation}

where $\mathcal{R}$ is the $n$--dimensional Ricci scalar, B.T. stands for boundary terms in order to have a well--defined action principle, $G_{n}$ the $n$--dimensional gravitational constant, $\alpha$ is a constant with units of length squared and $S_{m}$ is the action of the matter fields. The GB term $\mathcal{G}_{GB}$ reads as

\begin{equation}\label{GBinvariant}
\mathcal{G}_{\text{GB}}=R^2-4 R^{\mu \nu} R_{\mu \nu}+R^{\mu \nu \sigma \gamma} R_{\mu \nu \sigma \gamma}.
\end{equation}

It is worth mentioning that, in the case of superstring theory, in the low energy limit, $\alpha$ is related to the inverse string tension and is positive definite \cite{Gross:1986iv, Gross:1986mw}. Besides, the GB action is a natural extension of the Einstein theory in that no derivatives higher than second order appear in the field equations. For this to happen, the numerical coefficient of the $\mathcal{G}_{GB}$ must be $\{1,-4,1\}$ as shown in Eq. (\ref{GBinvariant}).

Now, taking variations with respect to metric tensor $g^{\mu\nu}$, the field equations coming from the action (\ref{fullaction}) are given by

\begin{equation}\label{fieldequations}
G^{\mu}_{\nu}+\alpha H^{\mu}_{\nu}=8 \pi G_{n} T^{\mu}_{\nu},
\end{equation}
being
\begin{equation}
    G_{\mu\nu}=R_{\mu\nu}-\frac{1}{2}g_{\mu\nu}\mathcal{R},
\end{equation}
the Einstein tensor and
             
\begin{align} \label{GBcontribution}
    H_{\mu \nu}&=2\left(\mathcal{R} R_{\mu \nu}-2 R_{\mu \lambda} R_\nu^\lambda-2 R^{\gamma \delta} R_{\gamma \mu \delta \nu}+R_\mu^{\alpha \gamma \delta} R_{\alpha \nu \gamma \delta}\right) \\& \nonumber -\frac{1}{2} g_{\mu \nu} \mathcal{G}_{G B}
\end{align}

the contribution coming from the GB term (\ref{GBinvariant}), the so--called Lanczos tensor. As we know, for $n=4$, the tensor $H_{\mu\nu}$ is zero because the GB term (\ref{GBinvariant}) becomes a topological invariant, not affecting the dynamic of the theory. 
Besides, $T^{\mu}_{\nu}$ is the energy--momentum tensor for matter fields obtained from $S_{m}$. 

Next, assuming that the energy-momentum tensor is the one describing an isotropic fluid
\begin{equation}
T_{\mu\nu}=\left(\rho+p\right)X_{\mu}X_{\nu}+pg_{\mu\nu},
\end{equation}
where $\rho$ is the density, $p$ the isotropic pressure and $X^{\mu}X_{\mu}=-1$ the four--velocity. The field Eqs. (\ref{fieldequations}) subject to the line element (\ref{OFLRW}) lead to the following Friedmann equations in the GB scenario
\begin{equation}\label{FGB1}
\frac{1}{R^{2}_{AH}}+\frac{\tilde{\alpha}}{R^{4}_{AH}}=\frac{16 \pi G_{n}}{(n-1)(n-2)} \rho,
\end{equation}
\begin{equation}\label{FGB2}
\left(1+2 \frac{\tilde{\alpha}}{R^{2}_{
AH}}\right)\left(\dot{H}-\frac{k}{a^2}\right)=-\frac{8 \pi G_{n}}{n-2}(\rho+p) .
\end{equation}
As expected, the above expressions are modified by the presence of the GB terms throughout the coupling constant $\tilde{\alpha}\equiv (n-3)(n-4)\alpha$ and also by the dimension $n$ of the space-time, recovering the usual Friedmann equations for the GR scenario when $n=4$.

In obtaining the above expressions, we have used Ricci's scalar
\begin{equation}
    \mathcal{R}=6\left(\frac{\dot{a}^{2}}{a^{2}}+\frac{\ddot{a}}{a}+\frac{k}{a^{2}}\right),
\end{equation}
and the GB term 
\begin{equation}
 \mathcal{G}_{\text{GB}}=24\left(\frac{k}{a^{2}}+H^{2}\right)\left(\dot{H}+H^{2}\right).
\end{equation}

\subsection{ UFL in GB gravity and Friedmann equations}\label{3SB}

In general, to deal with the thermodynamics of $AH$s, one needs to modify the usual approach based on the thermodynamics first law, employed in the static/stationary BHs case \cite{Bardeen:1973gs, Bekenstein:1973ur, Hawking:1975vcx, Gibbons:1976ue, York:1986it}. Hayward developed this approach in a series of articles \cite{Hayward:1993wb, Hayward:1994bu, Hayward:1997jp}. The main ingredients of this formulation are: i) the Misner--Sharp (MS) mass $M_{AH}$ \cite{Misner:1964je}, interpreted as the internal energy, ii) the HK surface gravity (\ref{SG}) (related with the temperature horizon) and iii) the areal volume $V_{AH}=\frac{4\pi}{3}R^{3}_{AH}$. Additionally, it is necessary to introduce new quantities to reach the desired thermodynamics description. These quantities are: iv) the density work $W$ and the v) energy--flux $\psi_{i}$ across the $AH$
\cite{Hayward:1997jp}. Using the metric in the representation given by Eq. (\ref{warped}), the definition of these new pieces is given by
\begin{equation}\label{workdensity}
    W\equiv-\frac{1}{2}h_{ij}T^{ij},
\end{equation}
and
\begin{equation}
    \psi_{i}\equiv T^{j}_{i}\nabla_{i}R+W\nabla_{i}R,
\end{equation}
respectively.

Putting together all these objects, the thermodynamics UFL acquires the following form \cite{Hayward:1993wb, Hayward:1994bu, Hayward:1997jp}
\begin{equation}
    \nabla_{i}M=A\psi_{i}+W\nabla_{i}V,
\end{equation}
or by identifying the mass $M$ with the internal energy $E$ one obtains 
\begin{equation}\label{UFL}
    \nabla_{i}E=A\psi_{i}+W\nabla_{i}V.
\end{equation}
It is worth mentioning that the term $A\psi_{i}$ is called the energy-supply vector.

As stated by Hayward in \cite{Hayward:1997jp}, the thermodynamics description comes when (\ref{UFL}) is projected along the direction of the $AH$, that is,
\begin{equation}\label{projected}  z^{i}\nabla_{i}E=\frac{\kappa_{\text{HK}}}{8\pi}z^{i}\nabla_{i}A+Wz^{i}\nabla_{i}V,
\end{equation}
where the following identification has been made $A\psi_{i}=\frac{\kappa_{\text{HK}}}{8\pi}z^{i}\nabla_{i}A$ (see \cite{Hayward:1997jp, Cai:2006rs} for further details about how to obtain this identification) and $z^{i}$ is a vector tangent to the $AH$. Interestingly, the Eq. (\ref{projected}) can be re-expressed as 
\begin{equation}\label{projected1}  
\frac{\kappa_{\text{HK}}}{8\pi}z^{i}\nabla_{i}A=R^{n-3}z^{i}\nabla_{i}\left(\frac{E}{R^{n-3}}\right)-Wz^{i}\nabla_{i}V,
\end{equation}
where it is argued that the first term on the right-hand member of the above expression becomes zero along the $AH$ \cite{Cai:2005ra, Helou:2015yqa}.

In \cite{Cai:2005ra, Akbar:2006kj, Cai:2006rs}, the above statement was employed (in the GR framework) to recognize the left-hand member as the Clausius relation. In this way, the second Friedmann equation is obtained using the corresponding entropy and the temperature identification with the HK surface gravity. Nevertheless, in \cite{Binetruy:2014ela, Helou:2015yqa, Helou:2015zma}, it was shown that these assumptions are not necessary to regain the second Friedmann equation. This is because the term claimed to be zero along the $AH$ is zero for any constant $R$ (including the particular case $R=R_{A}$). Moreover, projecting the remaining terms leads to the desired result without further consideration. The same argument was used in the context of GB gravity \cite{Cai:2005ra}. Nevertheless, the MS energy (mass) in the GB theory contains additional terms (corrections) \cite{Maeda:2006pm, Maeda:2007uu, Maeda:2008nz}, contributing to the equation of motion (see Eqs. (\ref{FGB1})--(\ref{FGB2})). Therefore, in the present case, it is not correct to argue that the first term in the right-hand side of Eq. (\ref{projected1}) is vanishing along the $AH$ because both the left-hand side and the second term in the right-hand side, are not containing any information about the corrections introduced by the GB term. The only way to assume the above is to set $AH$ to be constant and then assume the Clausius relation. However, this is a strong restriction because the $AH$ is dynamic. Then, to recast Friedmann Eqs. (\ref{FGB1})--(\ref{FGB2}) from the UFL (\ref{UFL}) we are going to project it along the Kodama vector direction (leading to Eq. (\ref{FGB2})) and along the direction orthogonal to the Kodama vector  (obtaining Eq. (\ref{FGB1})). It is worth mentioning that the information about the field equations is already contained in (\ref{UFL}). Therefore, by projecting it along different directions, we only isolate a particular information about the FLRW dynamic.

To proceed further, we must introduce an essential ingredient: the MS energy for GB theory. In \cite{Maeda:2006pm, Maeda:2007uu, Maeda:2008nz}, the spherically symmetric gravitational collapse of a dust cloud in the GB scenario, where the MS energy \cite{Misner:1964je} associated with this process was found to be

\begin{align}\label{MSEnergy}
E &\equiv \frac{(n-2)\pi^{(n-1)/2}}{8\pi G_{n}\Gamma\left[\frac{(n-1)}{2}\right]}\bigg[R^{n-3}\left(1-h^{ij}\nabla_{i}R\nabla_{j}R\right)\\&\nonumber+\tilde{\alpha} R^{n-5}\left(1-h^{ij}\nabla_{i}R\nabla_{j}R\right)^2\bigg].
\end{align}

The MS energy (\ref{MSEnergy}) evaluated at the FLRW metric (\ref{OFLRW}) reads
\begin{equation}\label{energyFLRW}
    E_{\text{FLRW}}=\frac{(n-2)\pi^{(n-1)/2}}{8\pi G_{n}\Gamma\left[\frac{(n-1)}{2}\right]}\frac{R^{n-1}}{R^{2}_{AH}}\left[1+\frac{\tilde{\alpha}}{R^{2}_{AH}}\right].
\end{equation}
Of course, the above expressions reduce to the GR for $n=4$.

It should be noted that the UFL (\ref{UFL}) was derived using the warped chart (\ref{warped}). As we are working in the Schwarzschild patch (\ref{schmetric}) where the time $T$ and radial $R$ coordinates are independent coordinates from each one, we need to transform every object of (\ref{UFL}) to get the desired result, taking into account that the area $A$, the work density $W$ and the volume $V$ do not transform since they are invariant (scalars). Therefore, for the component of the covariant derivative, one has 
\begin{eqnarray}\label{nabla1}
    \nabla_{T}&=&\frac{\partial t}{\partial T}\nabla_{t}+\frac{\partial r}{\partial T} \nabla_{r}=F \nabla_{t}-\frac{HRF}{a}\nabla_{r}, \\ \label{nabla2}
    \nabla_{R}&=& \frac{\partial t}{\partial R}\nabla_{t}+\frac{\partial r}{\partial R}\nabla_{r}= -\beta \nabla_{t}+\frac{1}{a}\nabla_{r},
\end{eqnarray}
while for the energy--flux vector 
\begin{eqnarray}\label{psi1}
    \psi_{T}&=&\frac{\partial t}{\partial T}\psi_{t}+\frac{\partial r}{\partial T} \psi_{r}=F \psi_{t}-\frac{HRF}{a}\psi_{r}, \\ \label{psi2}
    \psi_{R}&=& \frac{\partial t}{\partial R}\psi_{t}+\frac{\partial r}{\partial R}\psi_{r}= -\beta \psi_{t}+\frac{1}{a}\psi_{r},
\end{eqnarray}
obtaining 
\begin{eqnarray}\label{psi11}
    \psi_{T}&=&-HRF\left(p+\rho\right), \\ \label{psi22}
    \psi_{R}&=& \frac{1}{2}\left(1-k\frac{R^{2}}{a^{2}}\right)\left(1-\frac{R^{2}}{R^{2}_{AH}}\right)^{-1}\left(p+\rho\right).
\end{eqnarray}
To get (\ref{psi11}) and (\ref{psi22}) we used $\psi_{t}=-HR\left(p+\rho\right)/2$ and $\psi_{r}=a\left(p+\rho\right)/2$.

Now, the UFL (\ref{UFL}) projected on the Kodama (\ref{kodamasch}) vector yields to 
\begin{equation}  K^{T}\nabla_{T}E=AK^{T}\psi_{T}+WK^{T}\nabla_{T}V,
\end{equation}
giving
\begin{equation}\label{re1}
    \begin{split}
       K^{T}\nabla_{T}E=-\frac{(n-2)\pi^{(n-1)/2}}{4\pi G_{n}\Gamma\left[\frac{(n-1)}{2}\right]}F^{2}\sqrt{1-\frac{k}{a^{2}}R^{2}}\frac{R^{n-1}\dot{R}_{AH}}{R^{3}_{AH}}& \\ \times \left(1+2\frac{\tilde{\alpha}}{R^{2}_{AH}}\right), 
    \end{split}
\end{equation}
\begin{equation}\label{re2}
    WK^{T}\nabla_{T}V=0,
\end{equation}
\begin{equation}\label{re3}
    AK^{T}\psi_{T}=-\frac{2\pi^{(n-1)/2}}{\Gamma\left[\frac{(n-1)}{2}\right]}F^{2}HR^{n-1}\left(p+\rho\right).
\end{equation}
Putting together (\ref{re1}), (\ref{re2}) and (\ref{re3}) and after some algebra we obtain 
\begin{equation}\label{FGB21}
\left(1+2 \frac{\tilde{\alpha}}{R^{2}_{
AH}}\right)\left(\dot{H}-\frac{k}{a^2}\right)=-\frac{8 \pi G_{n}}{n-2}(\rho+p),
\end{equation}
where the following expression has been used in obtaining the above result
\begin{equation}
    \dot{R}_{AH}=-HR^{3}_{AH}\left(\dot{H}-\frac{k}{a^{2}}\right),
\end{equation}
recalling that every object is an implicit function of the $T$ time coordinate. Thus, the over-dot means differentiation with respect to $T$.

Next, the orthogonal vector to the Kodama vector is given by
\begin{equation}\label{orthogonal}
    M=M^{R}\partial_{R}=-\frac{1}{a}\sqrt{1-k\frac{R^{2}}{a^{2}}}\left(1-\frac{R^{2}}{R^{2}_{AH}}-H^{2}R^{2}\right)\partial_{R}.
\end{equation}
As can be seen, the vector (\ref{orthogonal}) is tangent to the hypersurface $\Sigma$, thus orthogonal to the Kodama vector (see Fig. \ref{fig1}), and hypersurface orthogonal to the two spheres defining the $AH$ (equivalently hypersurface orthogonal to the (n--2)--dimensional hypersurface (\ref{sphere})). However, it is not a null vector on (\ref{sphere}) as the Kodama vector does.

Projecting (\ref{UFL}) along the direction of (\ref{orthogonal}) we have
\begin{equation}\label{re11}
    \begin{split}
       M^{R}\nabla_{R}E=-\frac{(n-2)\pi^{(n-1)/2}}{8\pi a G_{n}\Gamma\left[\frac{(n-1)}{2}\right]}\sqrt{1-k\frac{R^{2}}{a^{2}}}& \\ \times\left[1-\frac{R^{2}}{R^{2}_{AH}}-H^{2}R^{2}\right]R^{n-2} &\\ \times \left[1-\frac{R^{2}}{R^{2}_{AH}}\right]^{-1}\bigg[-\left(n-1\right)\frac{H^{2}R^{n+2}}{R^{2}_{AH}}&\\ \times\left(1+\frac{\tilde{\alpha}}{R^{2}_{AH}}\right)+\frac{2H\dot{R}_{AH}R^{n+2}}{R^{3}_{AH}}&\\ \times\left(1+2\frac{\tilde{\alpha}}{R^{2}_{AH}}\right)+\frac{\left(n-1\right)}{R^{2}_{AH}}\left(1+\frac{\tilde{\alpha}}{R^{2}_{AH}}\right)&\\
       \times\left(1-\frac{R^{2}}{R^{2}_{AH}}\right)\bigg],
    \end{split}
\end{equation}
\begin{equation}\label{re22}
\begin{split}
    WM^{R}\nabla_{R}V=\frac{\pi^{(n-1)/2}}{\Gamma\left[(n-1)/2\right]}\left(1-\frac{R^{2}}{R^{2}_{AH}}\right)^{-1}&\\
    \times\left(1-\frac{R^{2}}{R^{2}_{AH}}-H^{2}R^{2}\right)\left(\rho-p\right)R^{n-2},
    \end{split}
\end{equation}
\begin{equation}\label{re33}
\begin{split}
    AM^{R}\psi_{R}=\frac{\pi^{(n-1)/2}}{\Gamma\left[(n-1)/2\right]}\left(1-\frac{R^{2}}{R^{2}_{AH}}\right)^{-1}&\\ \times
    \left(1-k\frac{R^{2}}{R^{2}_{AH}}\right)\left(\rho+p\right)R^{n-2}.
    \end{split}
\end{equation}
After some algebra, from expressions (\ref{re11}), (\ref{re22}) and (\ref{re33}) the output is
\begin{equation}\label{FGB11}
\frac{1}{R^{2}_{AH}}+\frac{\tilde{\alpha}}{R^{4}_{AH}}=\frac{16 \pi G_{n}}{(n-1)(n-2)} \rho.
\end{equation}
To obtain the first Friedmann Eq. (\ref{FGB11}) we have used $W=(\rho-p)/2$, $V=\pi^{(n-1)/2}R^{n-1}/\Gamma[(n+1)/2]$, $A=2\pi^{(n-1)/2}R^{n-2}/\Gamma[(n-1)/2]$ and the previous result (\ref{re22}) to cancel out the pressure $p$.

At this stage, some comments are pertinent. First of all, to obtain the Friedmann field equations (\ref{re22}) and (\ref{re11}), it is not necessary to assume from the very beginning the Clausius relation or include the associated entropy and temperature. Secondly, the complete information about the dynamic of the GB theory (in this case) in an FLRW background is already contained in the UFL (\ref{UFL}), and its projection on different directions isolates a specific part of it. As the Kodama vector defines a privileged time direction, it leads to a dynamical equation as (\ref{re22}), while along the orthogonal direction with respect to the Kodama flow, the vector (\ref{orthogonal}) provides the equation for the density $\rho$. Although we have not used the Clausius relation and the corresponding GB entropy and temperature to obtain the above result, it does not imply that this relation and associated quantities are not fundamental in the thermodynamical description of gravity \cite{Jacobson:1995ab}. In what follows, we are to get the GB entropy by identifying the surface gravity (\ref{SG}) with the temperature \cite{Bardeen:1973gs, Hawking:1975vcx, Bekenstein:1973ur, Hayward:1993wb, Hayward:1994bu, Hayward:1997jp, Jacobson:1995ab} and those terms accompanying it as the entropy variation. After integrating these terms, we will obtain the desired expression for the GB entropy. Of course, to achieve it, one needs to employ (\ref{UFL}) as the starting point in conjunction with the on-shell equations of motion. 

\subsection{The GB entropy}\label{3SC}

In the previous subsection \ref{3SB}, we obtained the dynamic description (equation of motion) for the FLRW space-time in the GB gravity theory. It was reached starting from the thermodynamics UFL (\ref{UFL}) and projecting it along different directions concerning the $AH$. In doing this, there was no assumption about temperature and entropy. Notwithstanding, as was proven and established at the starting works \cite{Bardeen:1973gs, Hawking:1975vcx, Bekenstein:1973ur, Gibbons:1976ue, York:1986it, Jacobson:1995ab}, there is a strong connection between the temperature and the surface gravity and also between the entropy and the area of the horizon (whatever the horizon, event or apparent horizons). Therefore, the link \emph{gravity} $\iff$ \emph{thermodynamics} becomes evident (see Fig. \ref{fig2} for a schematic representation of this conjecture). 

\begin{center}
\includegraphics[width=0.4\textwidth]{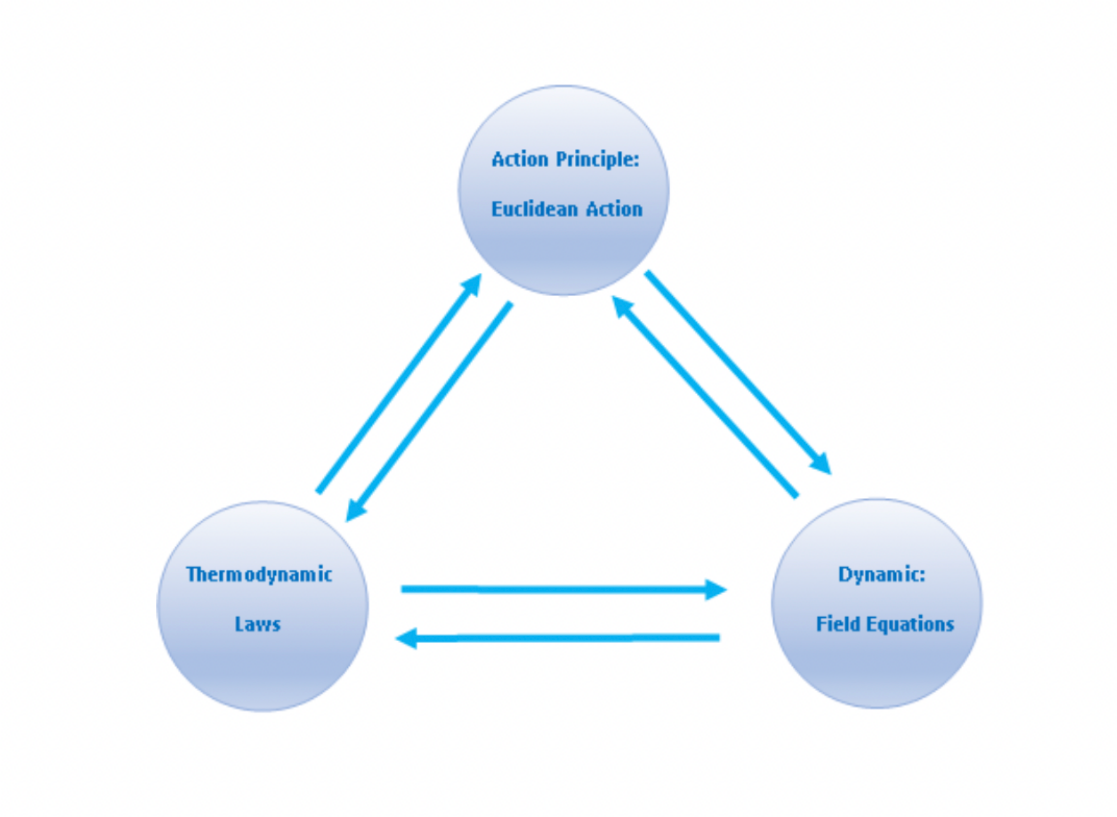}
    \captionof{figure}[111]{A schematic representation of the thermodynamics--gravity conjecture. Basically, this diagram reflects that, from the action principle, the field equations and thermodynamics description for a gravity theory are obtained. Furthermore, given a thermodynamics description, the field equations can also be obtained, and given the field equations, the thermodynamics description is reached, too.    }
    \label{fig2}
\end{center}

In this subsection, we want to obtain the entropy for the GB theory in the FLRW background. Supported by the bi-directional relationship between gravity and thermodynamics, we will start from the MS energy at the $AH$ together with the equations of motion. The differential form of the MS energy together with the UFL (\ref{UFL}) and the surface gravity (\ref{SG}) shall allow us to recognize the temperature and, consequently, the associated entropy. 
The result must be consistent with those expressions obtained for the GB entropy using the Euclidean path integral \cite{Gibbons:1976ue, York:1986it, Myers:1988ze} and also the so-called Wald entropy formula based on Noether charges \cite{Wald:1993nt, Clunan:2004tb}. It is worth mentioning that this approach has also been used in the context of BHs. These three ways to obtain the horizon entropy lead to the same expression. It can not be the exception for the cosmological scenario, independently of the gravitational theory.

The first assumption is to consider that the whole MS energy of the system is confined within an areal volume $V_{R_{AH}}$ along with a density $\rho_{AH}$, that is,
\begin{equation}\label{energyeval}
    E_{AH}=\rho_{AH}V_{AH}.
\end{equation}
The time variation of the above expression is
\begin{equation} 
\frac{dE_{AH}}{dt}=\rho_{AH}\frac{dV_{AH}}{dt}+V_{AH}\frac{d\rho_{AH}}{dt}.
\end{equation}
Now, making use of the work density $W$, the above equation can be recast as 
\begin{equation}\label{variationenergy}
\frac{dE_{AH}}{dt}=W\frac{dV_{AH}}{dt}+V_{AH}\frac{d\rho_{AH}}{dt}+\frac{\left(\rho_{AH}+p_{AH}\right)}{2}\frac{dV_{AH}}{dt}.
\end{equation}
The last two-term in the right-hand member of the previous expression provide 
\begin{equation}\label{resultvariation}
    \begin{split}
 V_{AH}\frac{d\rho_{AH}}{dt}+  \frac{\left(\rho_{AH}+p_{AH}\right)}{2}\frac{dV_{AH}}{dt}=\frac{(n-2)\pi^{(n-1)/2}}{2G_{n}\Gamma[(n-1)/2]}&\\ \times \left[1+2\frac{\tilde{\alpha}}{R^{2}_{AH}}\right] R^{n-3}_{AH}\frac{dR_{AH}}{dt} &\\ \times
 \left[\frac{1}{2\pi R_{AH}}\left(\frac{1}{2HR_{AH}}\frac{dR_{AH}}{dt}-1\right)\right].
    \end{split}
\end{equation}
Now, from (\ref{SG}), it is not hard to show that for the FLRW at the $AH$ one gets
\begin{equation}
    \kappa_{\text{HK}}=\frac{1}{ R_{AH}}\left(\frac{1}{2HR_{AH}}\frac{dR_{AH}}{dt}-1\right),
\end{equation}
and following \cite{Hawking:1975vcx,Hayward:1997jp} we know that
\begin{equation}\label{temperature}
    T\equiv\frac{\kappa_{\text{HK}}}{2\pi},
\end{equation}
from (\ref{resultvariation}), it is clear that the last term corresponds to the temperature of the system evaluated at the $AH$, so (\ref{resultvariation}) can be recast as
\begin{equation}\label{resultvariation1}
    \begin{split}
 V_{AH}\frac{d\rho_{AH}}{dt}+  \frac{\left(\rho_{AH}+p_{AH}\right)}{2}\frac{dV_{AH}}{dt}=\frac{(n-2)\pi^{(n-1)/2}}{2G_{n}\Gamma[(n-1)/2]}&\\ \times \left[1+2\frac{\tilde{\alpha}}{R^{2}_{AH}}\right] R^{n-3}_{AH}\frac{dR_{AH}}{dt} T_{AH}.
    \end{split}
\end{equation}

 Turning back to (\ref{variationenergy}) and plugging (\ref{resultvariation1}) we obtain 

\begin{equation}\label{variationenergy2}
\begin{split}
\frac{dE_{AH}}{dt}=W\frac{dV_{AH}}{dt}+\frac{(n-2)\pi^{(n-1)/2}}{2G_{n}\Gamma[(n-1)/2]}&\\ \times \left[1+2\frac{\tilde{\alpha}}{R^{2}_{AH}}\right] R^{n-3}_{AH}\frac{dR_{AH}}{dt} T_{AH}.
\end{split}
\end{equation}

So, as stated by Hayward in \cite{Hayward:1997jp}, once the UFL (\ref{UFL}) is projected (\ref{projected}) and the surface gravity related to the temperature, the variation of the area can be identified with the variation of the entropy, that is, $dA\sim dS$ or equivalently the energy--supply term is matching the Clausius relation. So, from  (\ref{variationenergy2}) we extract the following information for the GB entropy $S_{\text{GB}}$
\begin{equation}\label{entropy}
dS_{\text{GB}}=\frac{(n-2)\pi^{(n-1)/2}}{2G_{n}\Gamma\left[(n-1)/2\right]}\left[1+2\frac{\tilde{\alpha}}{R^{2}_{AH}}\right] {R}^{n-3}_{AH}dR_{AH},
\end{equation}
integrating the above expression it provides
\begin{equation}\label{entropy2}
    S_{\text{GB}}=\frac{A_{AH}}{4G_{n}}\left[1+2\frac{\tilde{\alpha}}{R^{2}_{AH}}\frac{(n-2)}{(n-4)}\right],
\end{equation}
or for a general areal radius $R$
\begin{equation}\label{entropy22}
    S_{\text{GB}}=\frac{A}{4G_{n}}\left[1+2\frac{\tilde{\alpha}}{R^{2}}\frac{(n-2)}{(n-4)}\right].
\end{equation}

This result (\ref{entropy2}) was previously obtained using the Euclidean Path integral in \cite{Myers:1988ze} and also using the Wald entropy formula in \cite{Clunan:2004tb}. As can be seen, all these approaches are consistent, yielding the same result. Despite the entropy (\ref{entropy2}) obtained for BHs solutions, this is also compatible with a cosmological scenario. The same happens in the GR framework for the FLRW cosmological model, where the entropy is $S=A/4$ as in the BH case \cite{Faraoni:2011hf}. Nevertheless, it should be considered that for the cosmological scenario, the area of the $AH$ is changing, leading to a dynamical entropy compared with its static/stationary BH counterpart where the area is constant. 
 
\section{Concluding remarks}\label{sec4}

The equations of motion, i.e., first and second Friedmann equations, for an FLRW model in the framework of GB gravity theory have been obtained, Eqs. (\ref{FGB21}) and (\ref{FGB11}). These equations were obtained by projecting the thermodynamics UFL (\ref{UFL}) along different directions with respect to the $AH$. For the second Friedmann equation (\ref{FGB21}), the UFL is projected along the direction of the Kodama vector field (\ref{kodamasch}). This Kodama vector establishes a privileged time direction (see Fig. \ref{fig1}). In the present context, this vector becomes a null vector on the $AH$. Then, the $AH$ behaves as a null surface. The Kodama vector is tangent to the hypersurface (\ref{sphere}) describing the $AH$. So, projecting the UFL (\ref{UFL}) along the Kodama vector field, thus along the $AH$, the acceleration equation is recovered. On the other hand, the projection of the UFL (\ref{UFL}) along the orthogonal direction of the Kodama vector, as (\ref{orthogonal}) does, the first Friedmann equation (\ref{FGB11}) is recast. Interestingly, the vector field (\ref{orthogonal}) is hypersurface orthogonal to the $AH$ (\ref{sphere}), then given the equation moderating the rate expansion of the Universe.

It is worth mentioning that, in deriving these equations, it is not necessary to assume the notion of temperature and entropy and, therefore, the Clausius relation. Nevertheless, it does not imply that Clausius's relation is not fundamental in the thermodynamics description of the actual scenario. Of course, as was pointed out by Hayward in \cite{Hayward:1997jp}, after projecting the UFL (\ref{UFL}), the energy--supply term can be interpreted as the heat flux (see the first term in the right member of Eq. (\ref{projected})). Thus, there is a correlation with the Clausius thermodynamics relation. The point is that, in projecting the UFL (\ref{UFL}) along the Kodama vector (\ref{kodamasch}), this term is vanishing, and the modifications to the Friedmann equations coming from the GB sector are introduced from the gradient of the MS energy member. Moreover, as we are dealing with a dynamical horizon, the $AH$, it is more natural to use the Kodama vector field (\ref{kodamasch}) to project the UFL (\ref{UFL}) along the $AH$ to isolated each Friedmann equation, already contained in the UFL (\ref{UFL}).

So, the previous result was obtained by applying the thermodynamics conjecture, starting from the thermodynamics description, to reach the dynamic equations of motion of the GB gravity theory in an FLRW background. Then, inversely applying the same methodology, from gravity to thermodynamics, we obtained the corrected or associated entropy to the GB theory (\ref{entropy2}). This result coincides with previous ones obtained using different (but at the same time equivalent) approaches, such as the Path integral \cite{Myers:1988ze} and Wald entropy formula \cite{Clunan:2004tb}. It should be noted that, in deriving (\ref{entropy2}), we have not considered any particular cosmological era, leading to an expanding, static, or contracting cosmology. This is so because, once a concrete era is considered, one needs to take into account the signature or global sign of the surface gravity (\ref{SG}) and temperature (\ref{temperature}). This is dictated by the $ AH$'s future (past) inner (outer) feature of the $AH$. Notwithstanding, the temperature expression appears on both sides and cancels out, leading without any extra information to the correct entropy (\ref{entropy2}).

So, in general, to apply the gravity-thermodynamics conjecture to cosmological scenarios, the relevant geometric object to do thermodynamics over the $ AH $ is the Kodama vector field. This is so because this object isolated the corresponding dynamic information without assuming strong restrictions on the entropy, temperature, and  Universe evolution. Besides, the inverse Path to regain thermodynamics potentials does not need to assume anything about the causal structure of the $AH$ to reach the desired and correct result.

\section*{ACKNOWLEDGEMENTS}
J. Saavedra and F. Tello-Ortiz acknowledge to grant FONDECYT N°1220065..
F. Tello-Ortiz acknowledges VRIEA-PUCV
for financial support through Proyecto Postdoctorado 2023 VRIEA-PUCV.

\bibliography{biblio.bib}
\bibliographystyle{elsarticle-num}

\end{document}